\documentclass[1p]{elsarticle}

\usepackage{csquotes}
\usepackage{amsmath}
\usepackage{amsfonts}
\usepackage{mathtools}
\usepackage{bbm}
\usepackage{amssymb}
\usepackage{graphicx}
\usepackage{epstopdf}
\usepackage{caption}
\usepackage{color}
\usepackage{subcaption}
\usepackage{siunitx}

\usepackage[mathlines]{lineno}

\journal{Physica A}

\DeclarePairedDelimiter\ceil{\lceil}{\rceil}

\begin{document}

\begin{frontmatter}

\title{Modeling random walkers on growing random networks}

\author[1]{Robert J. H. Ross\corref{cor1}}
\ead{robert\_ross@hms.harvard.edu}

\author[1]{Walter Fontana\corref{cor1}}
\ead{walter\_fontana@hms.harvard.edu}

\cortext[cor1]{Corresponding authors}

\address[1]{Department of Systems Biology, Harvard Medical School\\200 Longwood Avenue, Boston MA 02115}

\begin{abstract}
We present continuum models that describe the evolution of the position of a random walker on a growing network using four different growth algorithms. Three of these involve a random element, including one in which the motility rate of the random walker controls the network topology. For motility rates in which the position of the walker can be treated as quasi-stationary, we present accurate approximations to replace pair probabilities that allow us to numerically solve an otherwise intractable system of equations.
\end{abstract}


\end{frontmatter}

\section{Introduction}

Many systems of scientific interest can be modeled as dynamic processes situated on growing networks. These systems include technological structures such as the internet, social networks, and biological networks such as the vascular system and the mammalian brain \citep{Perra2012, Newman2010book, Dorogovtsev2013}.  The development of methods to model the behavior of dynamic processes situated on growing networks is still in progress however, as it presents numerous technical challenges \citep{Porter2016, Hoffmann2013, Holme2012}. In light of this, we demonstrate how to derive equations for the temporal evolution of probability distributions that describe the position of a random walker on networks whose growth process admits varying degrees of randomness. This work can be seen as an extension to previously presented results that describe the behavior of random walkers on growing lattices \citep{Ross2017a,Ross2016b,Ross2017b}. Continuum models often provide useful simplifications of complex stochastic systems affording insight not always apparent from studying simulation output alone. Our use of an unbiased random-walker as the process hosted by the network also has an intuitive physical interpretation applicable to a range of contexts, such as the diffusion of gases and heat, or migrating cells in development \citep{Reif,Redner2001,Hughes,Berg}. Furthermore, despite its relative simplicity, the random walker model can exhibit interesting behavior and can be viewed as a building block to generate more complex models.
 
The work we present here has two parts. We first present a discrete model of a random walker on a growing network that always remains fully connected and show how to derive equations describing the temporal evolution of the walker's position. Beginning with deterministic network growth allows us to introduce our approach in a simpler setting. We then present three distinct network growth algorithms that include random components. In two of these the position of the random walker on the network does not determine where the network grows. These two processes differ from each other in that for one growth process the expected node degree is constant, whereas in the other it is not. The third growth algorithm allows the position of the random walker on the network to determine where the network grows. In this instance, it has been previously demonstrated that the motility rate of the walker, $p_{m}$, determines the network structure \citep{Saramaki2004, Evans2005, Cannings2013, Ross2018b,Ross2018c}. For the three cases that allow for randomness in network structure we derive equations describing the temporal evolution of the degree of the node the random walker finds itself on. This provides an `inside' view of the evolving network structure. Our approach can be altered to account for the node position of the random walker instead. As is often the case, the master equations contain pair probabilities that prevent closure. For the case in which the random walker is fast relative to growth, and is therefore in quasi-equilibrium over the growing network, we present a simple and accurate approximation can be used to replace the pair probabilities in these equations. This closure is applicable to systems in which growth is slow compared to the evolution of a process situated on the network.

\section{Results}
\label{sec:results}

In all models, we denote the number of nodes in the network at time $t$ by $N(t)$, an integer, and the number of edges at time $t$ by $E(t)$, also an integer. Each node in the network is uniquely labelled by the number $i$ of the event that introduced it, $i \in \{ 1, 2, \ldots, N(t) \}$. Each node keeps the same label throughout a history (simulation). Thus, when a new node is added to the network its label is $N(t)$. The degree of a node in the network is denoted by $k$, and the degree of node $i$ is denoted by $k_{i}$. 

The initial network contains $N_{0}=N(0)$ nodes. In all of the network growth algorithms we consider, edges are undirected and unweighted, and there are no self-edges. The simulation of a model occurs in continuous time and proceeds according to the standard Monte-Carlo procedure (e.g.\@ \citep{Gillespie_orig}) wherein random walker movement and node addition are modeled as exponentially distributed events in a Markov chain. Attempted random walker movement events occur with rate $p_{m}$ per unit time. Stochastic network growth occurs by addition of a new (empty) node to the network at rate $p_{g}$ per unit time and is, therefore, linear. The specific models we consider differ in how a new node attaches to the network. Throughout this work we compare our equations describing how the position of a random walker evolves in time on a growing network with numerical ensemble averages from simulation. 

\subsection{A random walker on a growing fully connected network}
\label{sec:fcng}

We start out with network growth that is deterministic in the sense that there is no randomness in where the new node attaches to the network. (There is randomness in when a growth event happens.) This scenario offers a simple, introductory illustration for how to set up a master equation of a growing network. In this model, when a growth event occurs at time $t$, a single new node, $N(t)$, is added to the network and attached to all preexisting nodes, so that the network remains fully connected at all times. There is no limit to the number of nodes from which the network can be composed. As the network is fully connected, the random walker can move to any node from its current position. We refer to this model as `fully connected network growth' or FCNG. We provide an explanatory figure in the Supplementary Material (SM1). We next derive a probability master equation describing the position of the random walker. Although for simplicity we derive this equation by imagining a single random walker on the network, our results are applicable to scenarios where there are multiple random walkers on the network, as will become apparent.

To simplify notation, we denote by $N$ the size of the network at time $t$, thus removing explicit mention of $t$. We also assume that the infinitesimal duration $\delta t$ accommodates at most one event (movement or growth). In view of the FCNG mechanism, the probability $\rho_{N+1}(i;t+\delta t)$ that a random walker on a fully connected network of size $N+1$ at time $t+\delta t$ will be found at node $i \in \{1, 2, \ldots, N+1 \}$ is given by
\begin{align}
\begin{split}
\rho_{N+1}(i;t+\delta t) = & (1 - \delta t p_{g})\rho_{N+1}(i;t) + \delta t p_{g}\rho_{N}(i;t) - \delta t p_{m}\rho_{N+1}(i;t) \\
& + \delta t p_{m}\sum^{N+1}_{j \neq i}\dfrac{\rho_{N+1}(j;t)}{N},\qquad i=1,\ldots,N
\end{split}
\label{eq:gen2}
\\
\begin{split}
\rho_{N+1}(N+1; t+\delta t) = & (1 - \delta t p_{g})\rho_{N+1}(N+1; t) - \delta t p_{m}\rho_{N+1}(N+1; t) \\
& + \delta t p_{m}\sum^{N+1}_{j \neq N+1}\dfrac{\rho_{N+1}(j; t)}{N}.
\end{split}
\label{eq:genaux}
\end{align}
Equation \eqref{eq:gen2} describes the evolution of probabilities associated with nodes that existed in the $N$-sized network, and so are able to `inherit' probability mass from the $N$-sized network via network growth, as well as via the movement of the random walker on the network. Equation \eqref{eq:genaux} describes the evolution of probability associated with the new node $N+1$. On a network of size $N+1$ this node is not able to inherit probability mass from the $N$-sized network via network growth, and so only the movement of the random walker on the network can confer probability mass on to this new node.

We next sum \eqref{eq:gen2} and \eqref{eq:genaux} over all nodes in the network to obtain
\begin{align}
\sum^{N+1}_{i=1}\rho_{N+1}(i; t+\delta t) = 
(1 - \delta t p_{g})\sum^{N+1}_{i=1}\rho_{N+1}(i; t) + \delta t p_{g}\sum^{N}_{i=1}\rho_{N}(i; t),
\label{eq:eq3}
\end{align}
and make the following assumption:
\begin{align}
\sum^{N}_{j=1}\rho_{N}(j; t) = N\rho_{N}(i;t)=N\rho_{N}(t) \qquad \forall N.
\label{eq:ergodic}
\end{align}
This assumption is valid if $p_{m} \gg p_{g}$, which means the walker is essentially with equal probability at any node $i$ on the network.
Using assumption \eqref{eq:ergodic} in \eqref{eq:eq3}, re-indexing network size, dividing by $\delta t$ and taking the limit $\delta t \rightarrow 0$, yields
\begin{align}
\frac{\mathrm{d}\rho_{N}(t)}{\mathrm{d}t} = -p_{g}\rho_{N}(t) +  p_{g}\left(\dfrac{N-1}{N}\right)\rho_{N-1}(t).
\label{eq:eq5}
\end{align}
Equation \eqref{eq:eq5} is similar to one previously presented \citep{Ross2017a} in the context of a ring of nodes. It should be apparent that if we set the `dilution' term, $(N-1/N)$, to identity Eq. \eqref{eq:eq5} reduces to the evolution equations for the Poisson distribution describing how many growth events occur in a given time interval. Equation \eqref{eq:eq5} admits the solution
\begin{align}
\rho_{N}(t) = \rho_{N_0}(0) p_{g}^{(N-N_{0})}\left(\dfrac{N_{0}}{N}\right)\dfrac{t^{(N-N_{0})}}{(N-N_{0})!}\exp(-p_{g}t),
\label{eq:sol}
\end{align}
where $\rho_{N_0}(0)$ is the initial concentration of random walkers on the network. If we set $n = N - N_{0}$ in Eq. \eqref{eq:sol} then it reduces to the Poisson distribution multiplied by a prefactor.  This solution is appropriate in situations where Eq. \eqref{eq:ergodic} is a valid assumption as seen in Fig.\@ \ref{fig:figure1}. We provide further plots in the Supplementary material at a higher resolution (SM2).
\begin{figure}[!h]
\begin{center}
\includegraphics[width=0.6\textwidth]{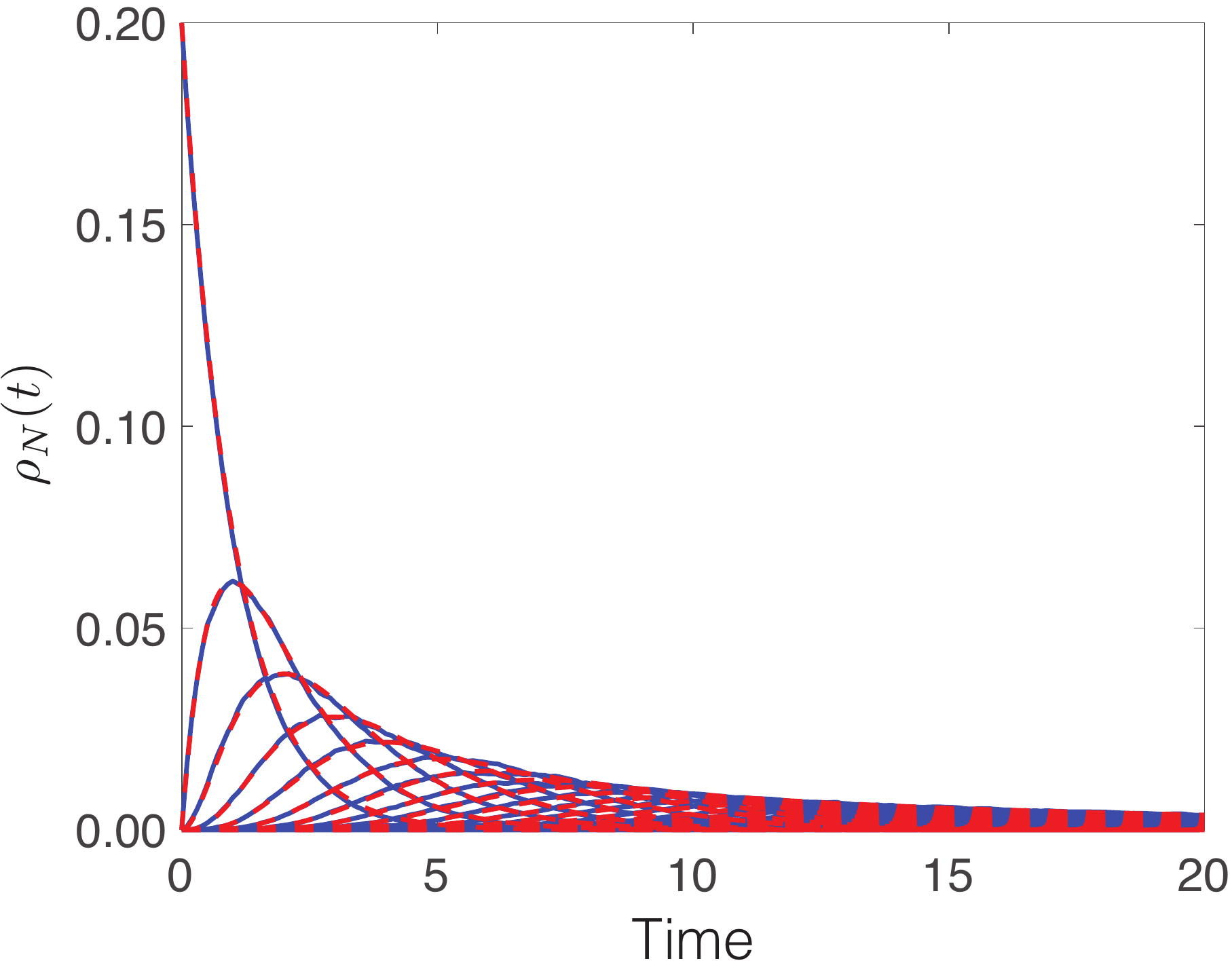}
\caption{Comparison of an ensemble average of the FCNG model and equation \eqref{eq:sol}. The red dashed lines refer to equation \eqref{eq:sol} for increasing values of $N$ (down and to the right). The blue lines report ensemble averages from simulation for network sizes $N=N_{0}+1, \ldots, N_{stop}$, with $N_{stop} = N_{0}+ \ceil{3p_{g}t_{end}}$ where $\ceil{\cdot}$ is the ceiling function and $t_{end}$ the simulated time. Ensemble averages are based on $10,000$ replicates of the discrete model. $N_{0}=5$, $p_{m}=10$ and $p_{g}=1$.}
\label{fig:figure1}
\end{center}
\end{figure}

\subsection{Random networks}

We next consider networks whose growth includes a random element. Our focus is on three growth algorithms.
In `random degree network growth' (RDNG), the degree $k$ of the new node is chosen at random up to current network size $N(t)$, and $k$ nodes from the network are chosen at random without replacement to link to the new node. In `random network growth' (RNG), the new node has degree $1$ and is linked up with one randomly chosen node of the network\footnote{Classically, RNG generates a random recursive tree.}. In `walker-induced network growth' (WING), the new node is linked up with the node at which the random walker is located. All schemes were introduced previously \citep{Dorogovtsev2013,Ross2018b,Ross2018c}, however, we provide explanatory figures for all three growth algorithms in the Supplementary Material (SM1). We first construct a \emph{general} master equation for network growth with randomness, into which terms specific to RDNG, RNG, or WING can be inserted. We then discuss this equation specifically in the context of WING. The derivations of the terms for RDNG and RNG are relegated to the Supplementary Material (SM3 and SM4). 

Since the network structure is no longer regular, as in FCNG, it is more meaningful to set up equations that describe the probability of the walker occupying a node of degree $k$ rather than an individual node $i$. However, the approach can be extended to include the node position of the random walker if so desired.

\subsubsection{General probability master equation}

We denote with $\rho_N([k]^1; t)$ the probability that a random walker occupies a node of degree $k$ on a growing (random) network of size $N$ at time $t$. $\rho_N([k]^1[j]^0; t)$ denotes the probability that the random walker is situated at a node of degree $k$ (indicated by the superscript $1$) and that a node of degree $k$ shares an edge with an unoccupied node of degree $j$ (indicated by the superscript $0$) in a network of size $N$ at time $t$. Throughout this section, the superscript `1' means a node is occupied by a random walker, and the superscript `0' means the node is not occupied by a random walker. We refer to probabilities of this kind as pair probabilities. Moreover, we introduce the term $\Gamma_{N}([k]^1)$ to mean the probability that, in a network of size $N$, a growth event attaches the new node to an occupied node of degree $k$. The term $\Gamma_{N}([k]^0)$ has the analogous meaning but for an unoccupied node of degree $k$. This covers growth mechanisms in which new nodes link up in ways that do not reference the location of the random walker. With this notation, the probability that a random walker occupies a node of degree $k$ on a growing random network of size $N$ at time $t+\delta t$ is given by
\begin{align}
\begin{split}
\rho_N([k]^1; t+\delta t) = \, & (1-\delta t p_g)\rho_N([k]^1; t)\\
& + \delta t p_g \Gamma_{N-1}([k-1]^1)\rho_{N-1}([k-1]^1; t)\\
& + \delta t p_g\left(1-\Gamma_{N-1}([k]^1)\right)\rho_{N-1}([k]^1; t)\\
& + \delta t p_m\sum_{j=1}^{N-1}\left(\alpha_k\rho_N([k]^0[j]^1; t)-\alpha_j\rho_N([k]^1[j]^0; t)\right).
\end{split}
\label{eq:irregEx_simp}
\end{align}
Here, $\alpha_{k}$ and $\alpha_{j}$ are constants that weigh the movement of the random walker between nodes of different degrees.

For FCNG we have
\begin{align}
\Gamma_{N-1}([k]^1) = \Gamma_{N-1}([k]^0) = 0 \quad\text{ and }\quad
\Gamma_{N-1}([k-1]^1) = \Gamma_{N-1}([k-1]^0) = 1.
\label{eq:gamma4}
\end{align}
Indeed, equations \eqref{eq:gamma4} reduce \eqref{eq:irregEx_simp} to the equations describing the position of a random walker on a growing fully connected network, \eqref{eq:gen2} and \eqref{eq:genaux}. 

The probability that a node of degree $k$ on a network of size $N$ is unoccupied at time $t+\delta t$ is 
\begin{align}
\begin{split}
\rho_N([k]^0; t+\delta t) = \, & (1-\delta t p_g)\rho_N([k]^0; t)\\
& + \delta t p_g \Gamma_{N-1}([k-1]^0)\rho_{N-1}([k-1]^0; t)\\
& + \delta t p_g\left(1-\Gamma_{N-1}([k]^0)\right)\rho_{N-1}([k]^0; t)\\
& + \delta t p_g\, [\text{source}]\\
& - \delta t p_m\sum_{j=1}^{N-1}\left(\alpha_k\rho_N([k]^0[j]^1; t)-\alpha_j\rho_N([k]^1[j]^0; t)\right).
\end{split}
\label{eq:irregEx_new_simp}
\end{align}
The `source' term in \eqref{eq:irregEx_new_simp} accounts for the rate at which empty nodes are being added to the network in growth events and depends on the growth mechanism. It will be specified shortly. To complete the equation for the evolution of $\rho_N([k]^1; t)$ we proceed as before by rearranging \eqref{eq:irregEx_simp} and \eqref{eq:irregEx_new_simp}, summing over the network size pertaining to each individual term\footnote{This is a conservation statement similar to that made in the derivation for FCNG. Intuitively, a node in a network of size $N$ has $N$ chances of being of degree $k$}, dividing by $\delta t$, and taking $\delta t \rightarrow 0$ in the limit to obtain
\begin{align}
\begin{split}
\dfrac{\mathrm{d}\rho_{N}([k]^1;t)}{\mathrm{d}t} = - & p_{g}\rho_{N}([k]^1;t) \\
& + \left(\dfrac{N-1}{N}\right)p_{g}\Gamma_{N-1}([k-1]^1)\rho_{N-1}([k-1]^1;t) \\
& + \left(\dfrac{N-1}{N}\right)p_{g}(1-\Gamma_{N-1}([k]^1))\rho_{N-1}([k]^1;t) \\
& + p_{m}\sum^{N-1}_{j=1}\left(\alpha_{k}\rho_N([k]^0[j]^1; t)-\alpha_j\rho_N([k]^1[j]^0; t)\right),
\end{split}
\label{eq:irregEx_simpD}
\end{align}
and
\begin{align}
\begin{split}
\dfrac{\mathrm{d}\rho_{N}([k]^0;t)}{\mathrm{d}t} = - & p_{g}\rho_{N}([k]^0;t) \\
& + \left(\dfrac{N-1}{N}\right)p_{g}\Gamma_{N-1}([k-1]^0)\rho_{N-1}([k-1]^0;t) \\
& + \left(\dfrac{N-1}{N}\right)p_{g}(1-\Gamma_{N-1}([k]^0))\rho_{N-1}([k]^0;t) \\
& + \dfrac{1}{N}p_g\, [\text{source}]\\
& + p_{m}\sum^{N-1}_{j=1}\left(\alpha_{k}\rho_N([k]^0[j]^1; t)-\alpha_j\rho_N([k]^1[j]^0; t)\right).
\end{split}
\label{eq:irregEx_new_simpD}
\end{align}

We next address the $\Gamma$ and source terms for RDNG, RNG and WING. In the Supplementary Material (SM3) we show that for RDNG
\begin{align}
\Gamma_{N-1}([k-1]^1) = \Gamma_{N-1}([k]^1) = \Gamma_{N-1}([k-1]^0) = \Gamma_{N-1}([k]^0) = \dfrac{N}{2(N-1)}.
\label{eq:gamUR1}
\end{align}
The source term for RDNG is
\begin{align}
[\text{source}]_{\text{RDNG}} = \dfrac{1}{N-1}\sum^{N-2}_{j=1}(\rho_{N-1}([j]^0;t) + \rho_{N-1}([j]^1;t)),\quad \text{for $k<N$}.
\label{eq:SourceUR1}
\end{align}
A new node is equally likely to be of any possible degree, hence the factor $1/(N-1)$. The addition of an empty node of degree $k$ is proportional to the probability that the network is of size $N-1$. We provide the full set of equations for RDNG in the Supplementary Material (SM4). It should be clear from the derivation of the RDNG master equations how to alter equations \eqref{eq:irregEx_simpD} and \eqref{eq:irregEx_new_simpD} to account for any desired network degree distribution.

In the case of RNG we have
\begin{align}
\Gamma_{N-1}([k-1]^1) = \Gamma_{N-1}([k-1]^0) = \dfrac{1}{N-1},
\label{eq:gamE1}
\end{align}
and
\begin{align}
\Gamma^{N-1}([k]^1) = \Gamma_{N-1}([k]^0) = \dfrac{N-2}{N-1},
\label{eq:gamE2}
\end{align}
with source term
\begin{align}
[\text{source}]_{\text{RNG}} = \sum^{N-2}_{j=1}(\rho_{N-1}([j]^0;t) + \rho_{N-1}([j]^1;t)),\quad \text{for $k=1$}.
\end{align}
In RNG, a new node is always of degree $k =1$. As in RDNG, the addition of an empty node of degree $k=1$ is proportional to the probability that the network is of size $N-1$. The full set of equations for RNG is laid out in the Supplementary material (SM5). 

Finally, in the case of WING, a node will only receive an edge from a new node if it is occupied by a walker:
\begin{align}
\Gamma_{N-1}([k-1]^1) = \Gamma_{N-1}([k]^1) = 1, \quad\text{ and }\quad \Gamma_{N-1}([k-1]^0) = \Gamma_{N-1}([k]^0) = 0.
\label{eq:gamPAD2}
\end{align}
The source term for WING is
\begin{align}
[\text{source}]_{\text{WING}} = \sum^{N-2}_{j=1}(\rho_{N-1}([j]^0;t) + \rho_{N-1}([j]^1;t)),\quad \text{for $k=1$},
\end{align}
and the full set of equations can be found in the Supplementary Material (SM6).

In Fig.\@ \ref{fig:figure2} we compare degree marginals of simulated ensemble averages with numerical solutions to the WING equations, \eqref{eq:irregEx_simpD} and \eqref{eq:irregEx_new_simpD} for $p_m=0$. This is also the case for RNG and RDNG, and shown in the Supplementary material (SM7).
\begin{figure}[!h]
\begin{center}
\includegraphics[width=1.04\textwidth]{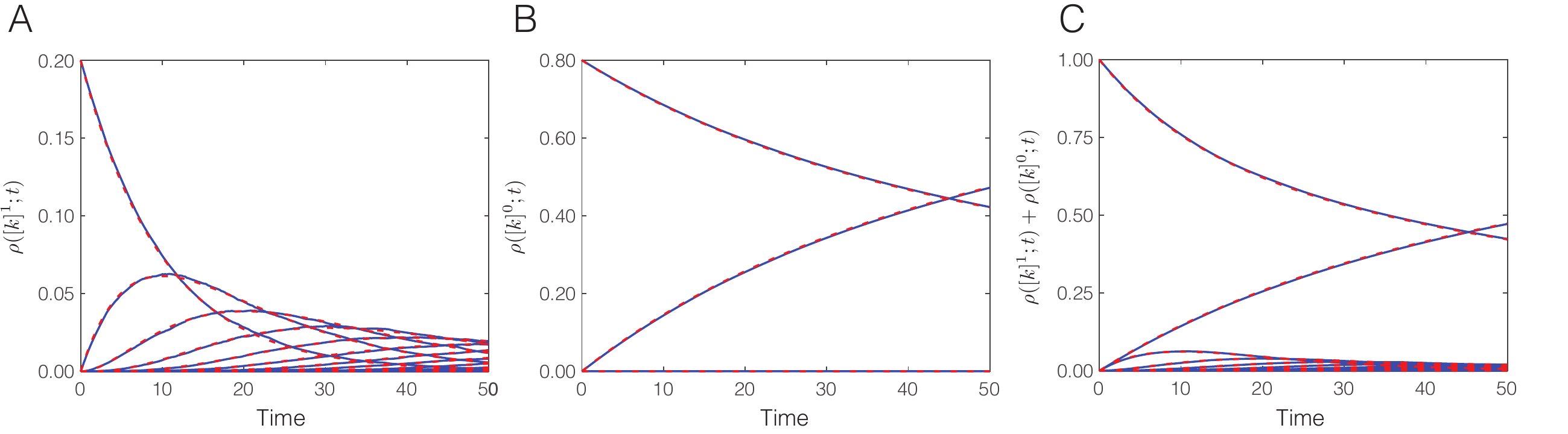}
\caption{Comparison of ensemble average and master equation for WING in the degree dimension. The blue lines are based on ensemble averages and plot $\rho([k]^1;t) = \sum^{N_{stop}-1}_{i=N_{0}}\rho_{i}([k]^1;t)$ and $\rho([k]^0;t) = \sum^{N_{stop}-1}_{i=N_{0}}\rho_{i}([k]^0;t)$ with $N_{stop}$ defined as in the caption of Fig.\@ \ref{fig:figure1}. The red dashed lines are based on numerical solutions of \eqref{eq:irregEx_simpD} and \eqref{eq:irregEx_new_simpD} for different values of $k$. Degree $k=N_{0}-1, N_{0}, \ldots, N_{stop}-1$ is increasing down and to the right in all panels. $N_{0}=5$, $p_{m}=0$ and $p_{g}=0.1$. Ensemble averages are over $10,000$ replicates of the simulated WING model. To solve the equations \eqref{eq:irregEx_simpD} and \eqref{eq:irregEx_new_simpD} we used MATLAB's \texttt{ode15s} for all network sizes up to $N_{stop}$.}
\label{fig:figure2}
\end{center}
\end{figure}

\begin{figure}[h!]
\begin{center}
\includegraphics[width=1\textwidth]{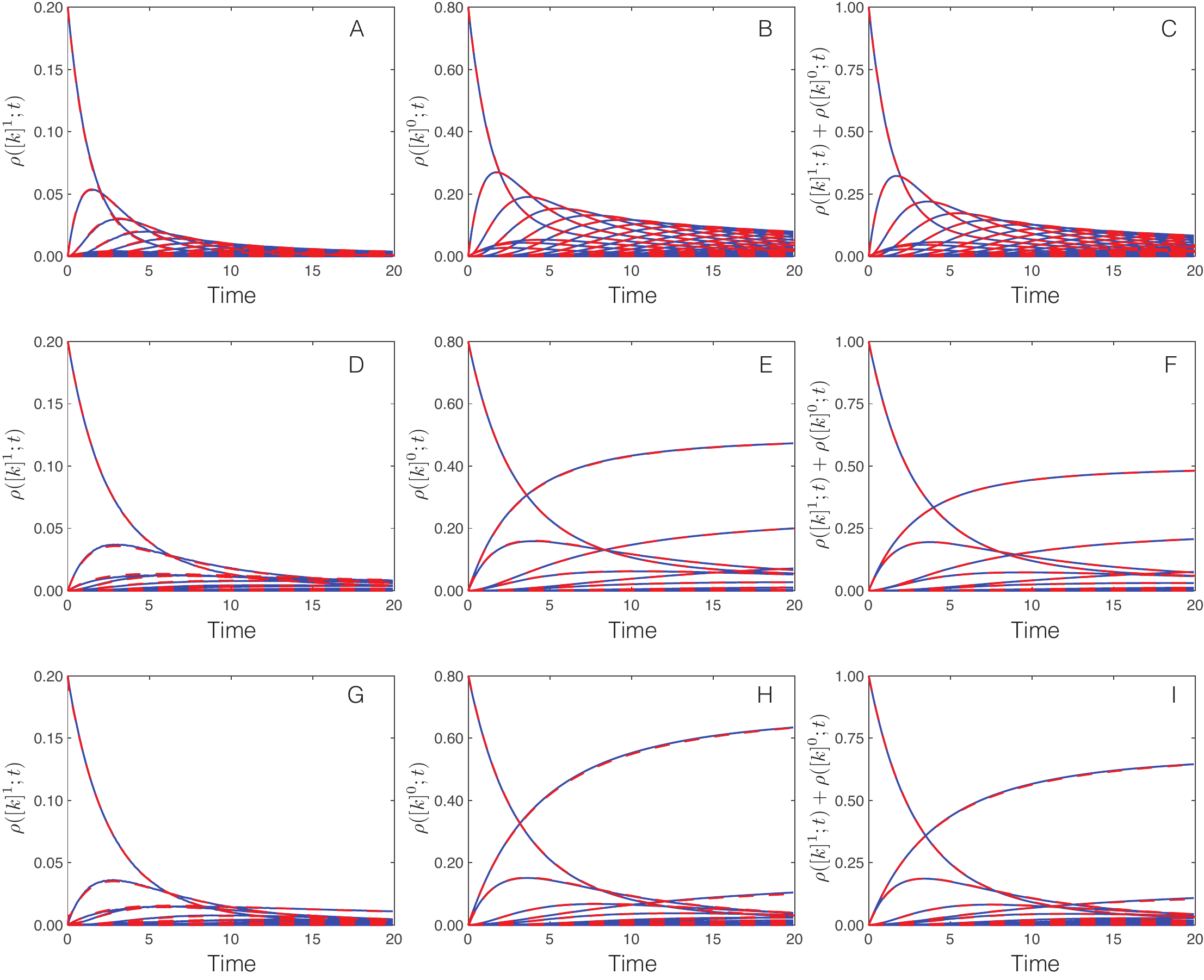}
\caption{Comparison of an ensemble average from various network growth models and the numerical solutions of \eqref{eq:irregEx_simpD} and \eqref{eq:irregEx_new_simpD} with appropriate source and $\Gamma$ terms and second order closure. Panels {\bf A}, {\bf B}, {\bf C}: RDNG; panels {\bf D}, {\bf E}, {\bf F}: RNG; panels {\bf G},{\bf H},{\bf I}: WING. The red dashed lines are numerical solutions of \eqref{eq:irregEx_simpD} and \eqref{eq:irregEx_new_simpD} for different values of $k$. The blue lines are ensemble averages from simulation. Degree $k=N_{0}-1, N_{0}, \ldots, N_{stop}-1$ increases down and to the right in all panels. Ensemble averages are based on $10,0000$ replicates. In all cases, $N_{0}=5$, $p_{g}=1$, and $p_{m} = 100$.}
\label{fig:figure3}
\end{center}
\end{figure}

In setting $p_{m} = 0$ we have removed the influence of the pair probabilities in equations \eqref{eq:irregEx_simpD} and \eqref{eq:irregEx_new_simpD}. For $p_m>0$, the pair probabilities evolve according to a set of equations that contain third order terms, which in turn refer to fourth order terms, and so on. However, if $p_m\gg p_g$, the position of the walker on the growing network can be treated as quasi-static (with respect to degree). In that case, the probability, $P_{w}$, that a walker is located at a given node is proportional to its in-degree $k$:
\begin{align}
P_{w}(k) \propto kP(k),
\label{eq:Pkprop}
\end{align}
where $P(k)$ is the degree distribution of the network. In this quasi-static case, we can therefore approximate the pairwise probabilities in the following manner:
\begin{align}
\alpha_{j}\rho_{N}([k]^1[j]^0;t) \approx \sum_{i=1}^{k}\frac{j}{k}\rho_{N}([k]^1;t)\rho_{N}([j]^0;t) = j\rho_{N}([k]^1;t)\rho_{N}([j]^0;t),
\label{eq:p2approxA}
\end{align}
and
\begin{align}
\alpha_{j}\rho_{N}([k]^0[j]^1;t) \approx \sum_{i=1}^{k}\frac{k}{j}\rho_{N}([k]^0;t)\rho_{N}([j]^1;t) = k\rho_{N}([k]^0;t)\rho_{N}([j]^1;t).
\label{eq:p2approx0}
\end{align}
Fig.\@ \ref{fig:figure3} shows the accuracy of this approximation when used in equations \eqref{eq:irregEx_simpD} and \eqref{eq:irregEx_new_simpD} for the various growth models. As $p_m\not\gg p_g$ the accuracy declines, most notably in the case of WING.  This is because in the case of WING as $p_{m}$ approaches $p_{g}$ the network structure becomes `stringy' and degree-degree correlations become significant \cite{Ross2018b}. In the Supplementary material (SM8) we demonstrate this breakdown in accuracy. For RNG and RDNG the network structure is not affected by $p_{m}$, and so this breakdown in accuracy does not occur to the same extent. To summarize, the closures in Eq. \eqref{eq:p2approxA} and \eqref{eq:p2approx0} are applicable when growth is slow compared to the evolution of a process situated on the network.

\section{Discussion}

We presented the construction and approximation of master equations for describing the movement of a random walker on growing networks based on four network growth algorithms. Three of these include a random element (RDNG, RNG, WING) and, among these, WING couples the location of growth to the position of the random walker. These growth procedures capture in a stylized manner different circumstances. For example, some networks grow in a fashion that is coupled to the behavior of a process situated on them, a scenario encapsulated by WING. Networks of this kind include the internet whose growth is determined by its usage and the developing brain whereby action potentials (a process situated on a neuronal network) help shape neuronal architecture (network structure). Given that growing spaces are central to phenomena of broad scientific interest \citep{Perra2012, Newman2010book, Dorogovtsev2013}, a concise description in terms of master equations can be of general use in studying them \citep{Ross2017b, Ross2018b, Ross2018c}.

We conclude by mentioning further research questions raised by this work.  An important problem is to find approximations for the pair probabilities that are accurate for all walker motility rates. This would widen the applicability of the master equations presented here to include systems in which the network growth rate is of a magnitude similar to the motility rate of the walker.  Doing so would be especially useful in the case of WING, in which the network structure is a function of the motility rate of the walker. Alternatively, evolution equations for the associated pair probabilities could be derived using the approach presented here, approximating the third order probabilities instead.  A large body of work is devoted to addressing pair probabilities in this way \citep{Hiebeler1997, Singer2004, Murrell2005, Murrell2010,Baker2010,Markham2013b,Markham2013a}. It would be interesting to understand whether such approaches could be deployed to study processes that are more complicated than the simple random walker model presented here, such as interactions among multiple random walkers or their proliferation.

\newpage

\section*{Supplementary Material}
\setcounter{section}{0}
\renewcommand{\thesection}{SM\the\value{section}}

\section{Figure explaining network growth mechanisms}
\label{sm:ngm}

\begin{figure}[h!]
\begin{center}
\includegraphics[width=1\textwidth]{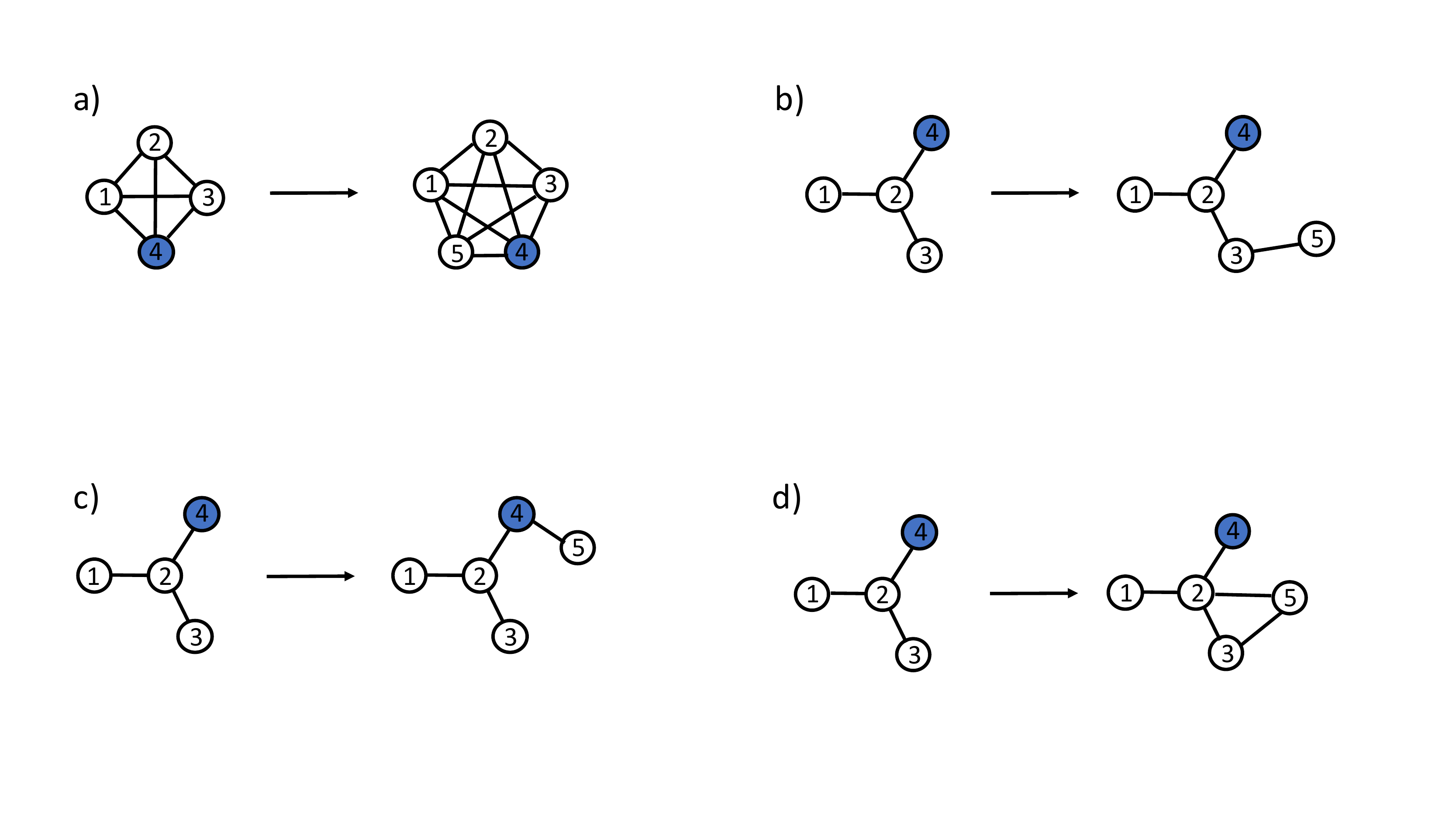}
\caption{The four network growth mechanisms we employ in this work. The position of the random walker on the network is denoted by blue, and the arrow indicates a growth event.  The random walker is able to move between any two nodes that share an edge. The nodes are labelled and keep the same label throughout a simulation replicate. (a) In FCNG all nodes in the network share an edge, and when a growth event occurs the new node creates a link to all pre-existing nodes in the network (b) In RNG when a growth event occurs the new node creates an edge with a node selected uniformly at random from the pre-existing nodes in the network. (c) In WING when a growth event occurs the new node creates an edge with the node upon which the random walker is currently situated. (d) In URNG when a growth event occurs the new node has degree $k$ selected uniformly at random from the set $\{1,2,...,N(t)\}$, and then $k$ nodes from the pre-existing network are chosen at random without replacement to link to the new node.}
\label{fig:figure4}
\end{center}
\end{figure}

\newpage
\section{Figures 1 and 3 (a)-(c) at a higher resolution}
\label{sm:zoom}

\begin{figure}[!h]
\centering
\includegraphics[width=1\textwidth]{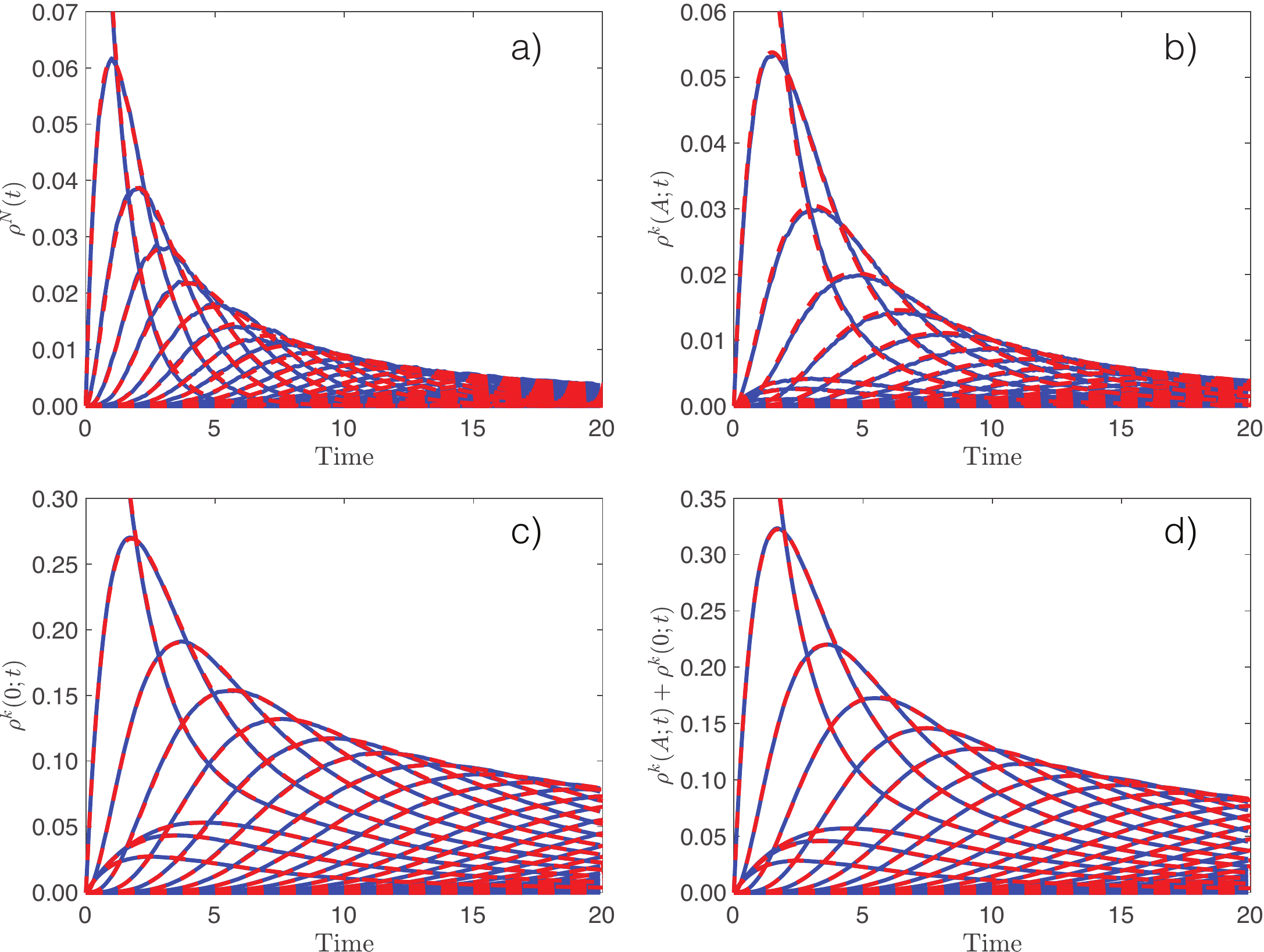}
\caption{Figures 1 and 3 (a)-(c) from the main text plotted with a different $y$-axis. (a) Figure 1, and (b)-(d) Figure 3 (a)-(c). All parameters are the same as given in the captions in the main text.}
\label{fig:figure5}
\end{figure}

\section{Derivation of $\Gamma$ terms for random degree network growth (RDNG)}
\label{sm:rdng}

We derive the probabilities $\Gamma_{N-1}([k-1]^1)$ and $\Gamma_{N-1}([k]^1)$ for RDNG.  As the position of the random walker on the network does not affect RDNG we know 
\begin{align}
\Gamma_{N-1}([k-1]^1) &= \Gamma_{N-1}([k-1]^0) = \Gamma_{N-1}(k-1),\\
\Gamma_{N-1}([k]^1) &= \Gamma_{N-1}([k]^0) = \Gamma_{N-1}(k).
\end{align}
Let $p(k)$ be the probability that the degree of a new node added to the network during a growth event is $k$, with the axiomatic constraint for $p(k)$ that
\begin{align}
\sum^{N-1}_{k=1} p(k) = 1.
\end{align}
The probability that an edge from the new node connects to a node of degree $k-1$ in a network of size $N-1$ during a growth event is
\begin{align}
\begin{split}
& \Gamma_{N-1}(k-1) = p(1)\frac{1}{N-1}+p(2)\left(\frac{1}{N-1}+\left(1-\frac{1}{N-1}\right)\frac{1}{N-2}\right) \\
& + p(3)\left(\frac{1}{N-1}+\left(1-\frac{1}{N-1}\right)\frac{1}{N-2}+\left(1-\frac{1}{N-1}\right)\left(1-\frac{1}{N-2}\right)\frac{1}{N-3}\right)\\
& +\ldots,
\end{split}
\label{eq:g1}
\end{align}
Equation \eqref{eq:g1} can be simplified to obtain
\begin{align}
\begin{split}
\Gamma_{N-1}(k-1) &= \frac{1}{N-1}\sum^{N-1}_{n=1}p(k) + \frac{1}{N-2}\sum^{N-1}_{n=2}p(k)\left(1-\frac{1}{N-1}\right)\\
& + \frac{1}{N-3}\sum^{N-1}_{n=3}p(k)\left(1-\frac{1}{N-1}\right)\left(1-\frac{1}{N-2}\right) + \ldots,
\end{split}
\end{align}
and further simplified
\begin{align}
\begin{split}
\Gamma_{N-1}(k-1) &= \frac{1}{N-1}\sum^{N-1}_{n=1}p(k) + 
\frac{1}{N-2}\sum^{N-1}_{n=2}p(k)\frac{N-2}{N-1} \\
& + \frac{1}{N-3}\sum^{N-1}_{n=3}p(k)\frac{N-3}{N-1} + \ldots, 
\end{split}
\end{align}
to arrive at
\begin{align}
\Gamma_{N-1}(k-1) = \frac{1}{N-1}\sum_{m=1}^{N-1}\sum_{n=m}^{N-1}p(k).
\label{eq:ERgamma}
\end{align}
Following the same reasoning we obtain
\begin{align}
\Gamma_{N-1}(k) = \frac{1}{N-1}\sum_{m=1}^{N-1}\sum_{n=m}^{N-1}p(k). \nonumber
\label{}
\end{align}
If $p(k) = 1/(N-1)$, that is, the degree of the new node is selected uniformly at random from $k \in \{1, \ 2, ..., \ N(t)\}$, then \eqref{eq:ERgamma} becomes
\begin{align}
\Gamma_{N-1}(k-1) = \frac{1}{N-1}\sum_{m=1}^{N-1}\sum_{n=m}^{N-1}\frac{1}{N-1} = \frac{1}{(N-1)^{2}}\frac{N(N-1)}{2}= \frac{N}{2(N-1)}.
\label{eq:gamma}
\end{align}

\section{RDNG probability master equation}
\label{sm:rdng_master}

The equations describing the evolution of $\rho_{N}([k]^1;t)$ when $1 < k < N-1$ take the form
\begin{align}
\begin{split}
\frac{\mathrm{d}\rho_{N}([k]^1;t)}{\mathrm{d}t} &= -p_g\rho_{N}([k]^1;t) \\
&  + p_g\frac{1}{2} \rho_{N-1}([k-1]^1;t) \\
&  + p_g\frac{N-2}{2N} \rho_{N-1}([k]^1;t) \\
&  + p_m\sum^{N-1}_{j=1}\left(\alpha_{k}\rho_{N}([k]^0[j]^1;t) - \alpha_{j}\rho_{N}([k]^1[j]^0;t)\right),
\end{split}
\end{align}
while for $k = 1$ and $k = N-1$ we have
\begin{align}
\begin{split}
\frac{\mathrm{d}\rho_{N}([1]^1;t)}{\mathrm{d}t} &= -p_g\rho_{N}([1]^1;t) \\
& + p_g\frac{N-2}{2N} \rho_{N-1}([1]^1;t) \\
& + p_m\sum^{N-1}_{j=1}\left(\alpha_{1}\rho_{N}([1]^0[j]^1;t) - \alpha_{j}\rho_{N}([1]^1[j]^0;t)\right),
\end{split}
\end{align}
and
\begin{align}
\begin{split}
\frac{\mathrm{d}\rho_{N}([N-1]^1;t)}{\mathrm{d}t} &= -p_g\rho_{N}([N-1]^1;t) \\
& + p_g\frac{1}{2} \rho_{N-1}([N-2]^1;t) \\
& + p_m\sum^{N-1}_{j=1}\left(\alpha_{N-1}\rho_{N}([N-1]^0[j]^1;t) - \alpha_{j}\rho_{N}([N-1]^1[j]^0;t)\right),
\end{split}
\end{align}
respectively.
The system of equations describing the evolution of $\rho_{N}([k]^0;t)$ when $1 < k < N-1$ takes the form
\begin{align}
\begin{split}
\frac{\mathrm{d}\rho_{N}([k]^0;t)}{\mathrm{d}t} &= -p_g\rho_{N}([k]^0;t) \\
& + p_g\frac{1}{2} \rho_{N-1}([k-1]^0;t) \\
& + p_g\frac{N-2}{2N} \rho_{N-1}([k]^0;t) \\
& + \frac{1}{N}\frac{p_g}{N-1}\sum^{N-2}_{j=1}\left(\rho_{N-1}([j]^0;t) + \rho_{N-1}([j]^1;t)\right)\\
& - p_m\sum^{N-1}_{j=1}\left(\alpha_{k}\rho_{N}([k]^0[j]^1;t) - \alpha_{j}\rho_{N}([k]^1[j]^0;t)\right),
\end{split}
\end{align}
while for $k = 1$ and $k = N-1$ we have
\begin{align}
\begin{split}
\frac{\mathrm{d}\rho_{N}([1]^0;t)}{\mathrm{d}t} &= -p_g\rho_{N}([1]^0;t) \\
& + p_g\frac{N-2}{2N} \rho_{N-1}([1]^0;t) \\
& + \frac{1}{N}\frac{p_g}{N-1}\sum^{N-2}_{j=1}\left(\rho_{N-1}([j]^0;t) + \rho_{N-1}([j]^1;t)\right)\\
& - p_m\sum^{N-1}_{j=1}\left(\alpha_{1}\rho_{N}([1]^0[j]^1;t) - \alpha_{j}\rho_{N}([1]^1[j]^0;t)\right),
\end{split}
\end{align}
and
\begin{align}
\begin{split}
\frac{\mathrm{d}\rho_{N}([N-1]^0;t)}{\mathrm{d}t} &= -p_g\rho_{N}([N-1]^0;t) \\
& + p_g\frac{1}{2} \rho_{N-1}([N-2]^0;t) \\
& + \frac{1}{N}\frac{p_g}{N-1}\sum^{N-2}_{j=1}\left(\rho_{N-1}([j]^0;t) + \rho_{N-1}([j]^1;t)\right)\\
& - p_m\sum^{N-1}_{j=1}\left(\alpha_{N-1}\rho_{N}([N-1]^0[j]^1;t) - \alpha_{j}\rho_{N}([N-1]^1[j]^0;t)\right),
\end{split}
\end{align}
respectively.

\section{Random network growth (RNG) probability master equation}
\label{sm:rng_master}

The equations describing the evolution of $\rho_{N}([k]^1;t)$ when $1 < k < N-1$ are
\begin{align}
\begin{split}
\frac{\mathrm{d}\rho_{N}([k]^1;t)}{\mathrm{d}t} &= -p_g\rho_{N}([k]^1;t) \\
& + p_g \frac{N-1}{N} \rho_{N-1}([k-1]^1;t) \\
& + p_m\sum^{N-1}_{j=1}\left(\alpha_{k}\rho_{N}([k]^0[j]^1;t) - \alpha_{j}\rho_{N}([k]^1[j]^0;t)\right),
\end{split}
\label{eq:irregEx_langE_simp_gamma21}
\end{align}
while for $k = 1$ and $k = N-1$ we have
\begin{align}
\begin{split}
\frac{\mathrm{d}\rho_{N}([1]^1;t)}{\mathrm{d}t} &= -p_g\rho_{N}([1]^1;t) \\
& + p_m\sum^{N-1}_{j=1}\left(\alpha_{1}\rho_{N}([1]^0[j]^1;t) - \alpha_{j}\rho_{N}([1]^1[j]^0;t)\right),
\end{split}
\label{eq:irregEx_langE_simp_gamma22}
\end{align}
and
\begin{align}
\begin{split}
\frac{\mathrm{d}\rho_{N}([N-1]^1;t)}{\mathrm{d}t} &= -p_g\rho_{N}([N-1]^1;t) \\
& + p_g \frac{N-1}{N} \rho_{N-1}([N-2]^1;t) \\
& + p_m\sum^{N-1}_{j=1}\left(\alpha_{N-1}\rho_{N}([N-1]^0[j]^1;t) - \alpha_{j}\rho_{N}([N-1]^1[j]^0;t)\right),
\end{split}
\label{eq:irregEx_langE_simp_gamma23}
\end{align}
respectively. The system of equations describing the evolution of $\rho_{N}([k]^0;t)$ when $1 < k < N-1$ take the form
\begin{align}
\begin{split}
\frac{\mathrm{d}\rho_{N}([k]^0;t)}{\mathrm{d}t} &= -p_g\rho_{N}([k]^0;t)\\
& + p_g\frac{1}{2} \rho_{N-1}([k-1]^0;t) \\
& + p_g\frac{N-2}{2N} \rho_{N-1}([k]^0;t) \\
& + \frac{1}{N}\frac{p_g}{N-1} \sum^{N-2}_{j=1}\left(\rho_{N-1}([j]^0;t) + \rho_{N-1}([j]^1;t)\right) \\
& - p_m\sum^{N-1}_{j=1}\left(\alpha_{k}\rho_{N}([k]^0[j]^1;t) - \alpha_{j}\rho_{N}([k]^1[j]^0;t)\right),
\end{split}
\label{eq:irregEx_langE_simp_gamma4}
\end{align}
while for $k = 1$ and $k = N-1$ we have
\begin{align}
\begin{split}
\frac{\mathrm{d}\rho_{N}([1]^0;t)}{\mathrm{d}t} &= -p_g\rho_{N}([1]^0;t) \\
& + p_g\frac{N-2}{2N} \rho_{N-1}([1]^0;t) \\
& + \frac{1}{N}\frac{p_g}{N-1} \sum^{N-2}_{j=1}\left(\rho_{N-1}([j]^0;t) + \rho_{N-1}([j]^1;t)\right) \\
& - p_m\sum^{N-1}_{j=1}\left(\alpha_{1}\rho_{N}([1]^0[j]^1;t) - \alpha_{j}\rho_{N}([1]^1[j]^0;t)\right),
\end{split}
\label{eq:irregEx_langE_simp_gamma5}
\end{align}
and
\begin{align}
\begin{split}
\frac{\mathrm{d}\rho_{N}([N-1]^0;t)}{\mathrm{d}t} &= -p_g\rho_{N}([N-1]^0;t) \\
& + p_g\frac{1}{2} \rho_{N-1}([N-2]^0;t) \\
& + \frac{1}{N}\frac{p_g}{N-1}\sum^{N-2}_{j=1}\left(\rho_{N-1}([j]^0;t) + \rho_{N-1}([j]^1;t)\right)\\
& - p_m\sum^{N-1}_{j=1}\left(\alpha_{N-1}\rho_{N}([N-1]^0[j]^1;t) - \alpha_{j}\rho_{N}([N-1]^1[j]^0;t)\right),
\end{split}
\end{align}
respectively. 

\section{Walker-induced network growth (WING) probability master equation}
\label{sm:wing_master}

The equations describing the evolution of $\rho_{N}([k]^1;t)$ for WING when $1 < k < N-1$ are
\begin{align}
\begin{split}
\frac{\mathrm{d}\rho_{N}([k]^1;t)}{\mathrm{d}t} &= -p_g\rho_{N}([k]^1;t) \\
& + p_g\frac{N-1}{N} \rho_{N-1}([k-1]^1;t) \\
& + p_m\sum^{N-1}_{j=1}\left(\alpha_{k}\rho_{N}([k]^0[j]^1;t) - \alpha_{j}\rho_{N}([k]^1[j]^0;t)\right),
\end{split}
\label{eq:irregEx_lang_simp_gamma21}
\end{align}
while for $k = 1$ and $k = N-1$ we have
\begin{align}
\begin{split}
\frac{\mathrm{d}\rho_{N}([1]^1;t)}{\mathrm{d}t} &= -p_g\rho_{N}([1]^1;t) \\
& + p_m\sum^{N-1}_{j=1}\left(\alpha_{1}\rho_{N}([1]^0[j]^1;t) - \alpha_{j}\rho_{N}([1]^1[j]^0;t)\right),
\end{split}
\label{eq:irregEx_lang_simp_gamma22}
\end{align}
and
\begin{align}
\begin{split}
\frac{\mathrm{d}\rho_{N}([N-1]^1;t)}{\mathrm{d}t} &= -p_g\rho_{N}([N-1]^1;t)  \\
&  + p_g \frac{N-1}{N}\rho_{N-1}([N-2]^1;t)  \\
&  + p_m\sum^{N-1}_{j=1}\left(\alpha_{N-1}\rho_{N}([N-1]^0[j]^1;t) - \alpha_{j}\rho_{N}([N-1]^1[j]^0;t)\right),
\end{split}
\label{eq:irregEx_lang_simp_gamma23}
\end{align}
respectively. The system of equations describing the evolution of $\rho_{N}([k]^0;t)$ for WING are
\begin{align}
\begin{split}
\frac{\mathrm{d}\rho_{N}([k]^0;t)}{\mathrm{d}t} &= -p_g\rho_{N}([k]^0;t) \\
& + p_g \frac{N-1}{N}\rho_{N-1}([k]^0;t) \\
& - p_m\sum^{N-1}_{j=1}\left(\alpha_{k}\rho_{N}([k]^0[j]^1;t) - \alpha_{j}\rho_{N}([k]^1[j]^0;t)\right),
\end{split}
\label{eq:irregEx_lang_simp_gamma24}
\end{align}
while for $k = 1$ and $k = N-1$ we have
\begin{align}
\begin{split}
\frac{\mathrm{d}\rho_{N}([1]^0;t)}{\mathrm{d}t} &= -p_g\rho_{N}([1]^0;t)  \\
& + p_g\frac{N-1}{N}\rho_{N-1}([1]^0;t)  \\
& + frac{1}{N} p_g \sum^{N-2}_{j=1}\left(\rho_{N-1}([j]^0;t) + \rho_{N-1}([j]^1;t)\right)  \\
& - p_m\sum^{N-1}_{j=1}\left(\alpha_{1}\rho_{N}([1]^0[j]^1;t) - \alpha_{j}\rho_{N}([1]^1[j]^0;t)\right),
\end{split}
\label{eq:irregEx_lang_simp_gamma25}
\end{align}
and
\begin{align}
\begin{split}
\frac{\mathrm{d}\rho_{N}([N-1]^0;t)}{\mathrm{d}t} &= -p_g\rho_{N}([N-1]^0;t) \\
& - p_m\sum^{N-1}_{j=1}\left(\alpha_{N-1}\rho_{N}([N-1]^0[j]^1;t) - \alpha_{j}\rho_{N}([N-1]^1[j]^0;t)\right),
\end{split}
\label{eq:irregEx_lang_simp_gamma26}
\end{align}
respectively. 

\section{RDNG when $p_m=0$}
\label{sm:rdng_pmzero}

\begin{figure}[!h]
\begin{center}
\includegraphics[width=1\textwidth]{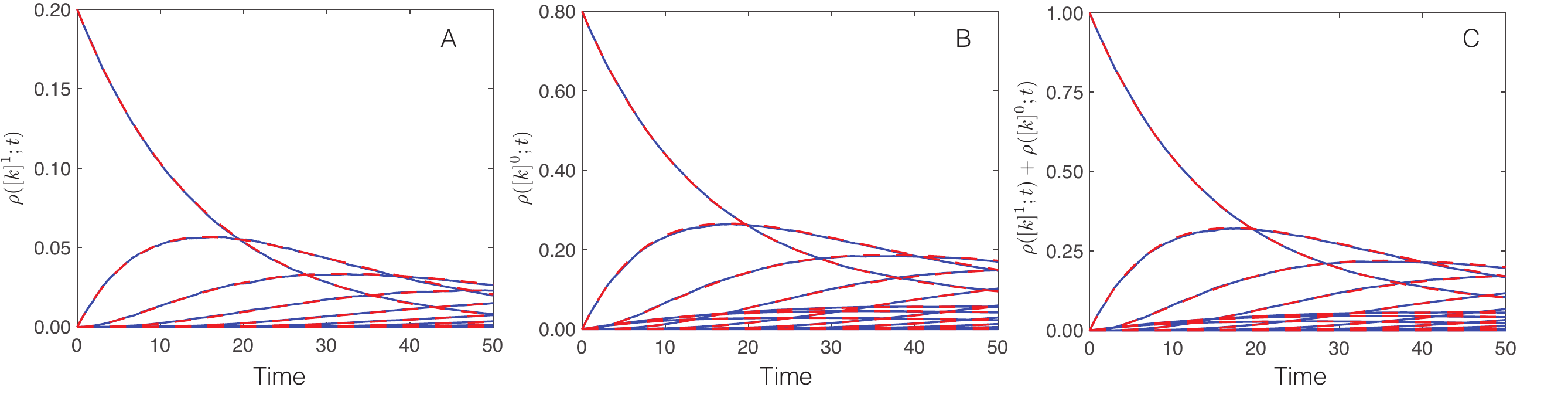}
\caption{Comparison of ensemble averages for RDNG and numerical solutions of equations \eqref{eq:irregEx_simpD} and \eqref{eq:irregEx_new_simpD} in the main text. The red dashed lines are the numerical solutions of \eqref{eq:irregEx_simpD} and \eqref{eq:irregEx_new_simpD} for different values of $k$. The blue lines are the corresponding ensemble averages. Increasing degree $k=-N_{0}-1,N_{0}, \ldots, N_{stop}-1$ is down and to the right in all panels. Ensemble averages are taken from $10,000$ replicates. $N_{0}=5$, $p_m=0$ and $p_g=0.1$}
\label{fig:figure6}
\end{center}
\end{figure}

\newpage
\section{Validity of quasi-static approximation when $p_m = p_{g}$}
\label{sm:wing_breakdown}

\begin{figure}[!h]
\begin{center}
\includegraphics[width=1\textwidth]{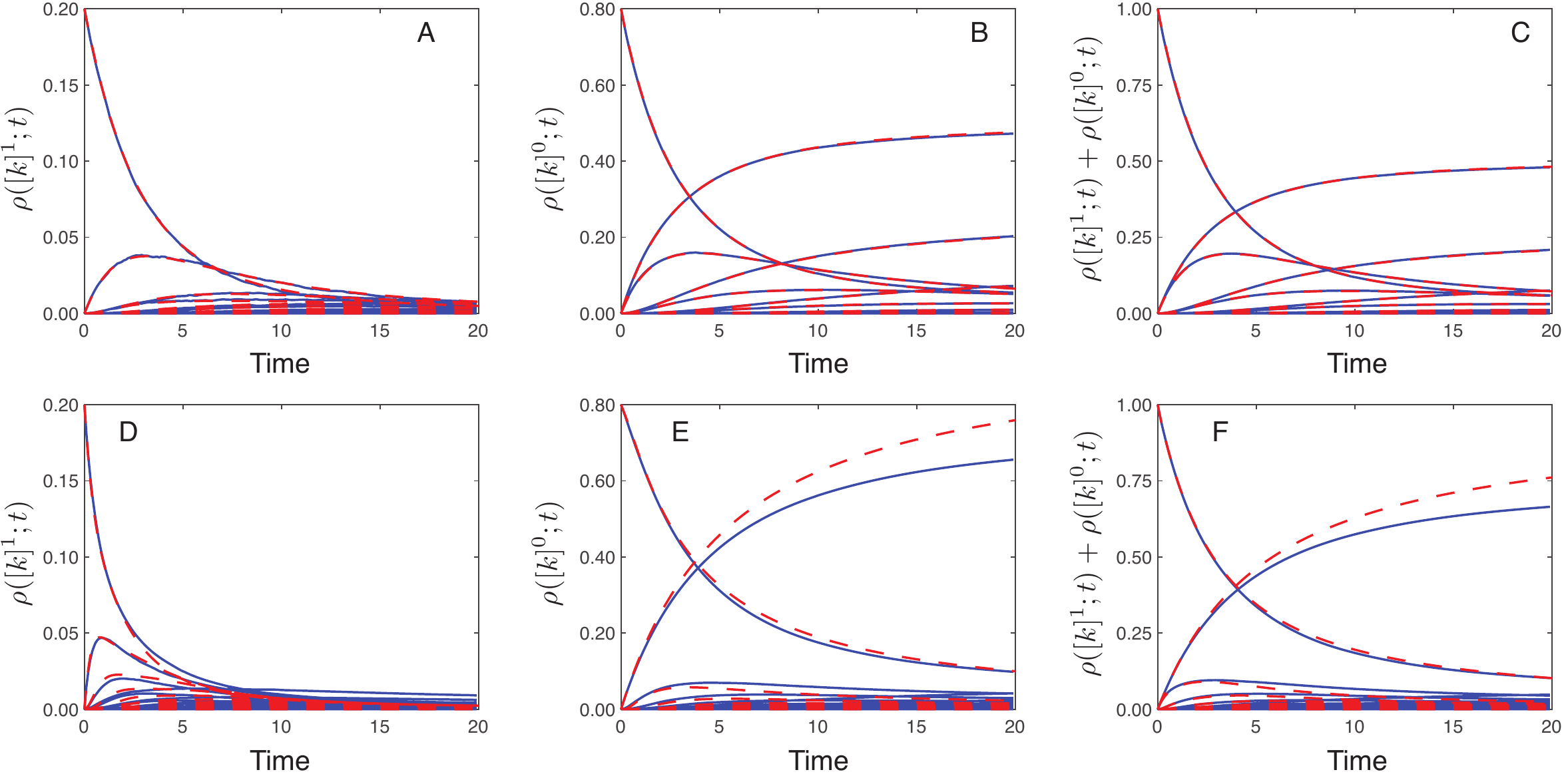}
\caption{Comparison of an ensemble average for RNG and WING and the numerical solutions of equations \eqref{eq:irregEx_simpD} and \eqref{eq:irregEx_new_simpD} with appropriate source and $\Gamma$ terms and second order closure. Panels {\bf A}, {\bf B}, {\bf C}: RNG; panels {\bf D}, {\bf E}, {\bf F}: WING. The red dashed lines are numerical solutions of \eqref{eq:irregEx_simpD} and \eqref{eq:irregEx_new_simpD} for different values of degree $k=N_{0}-1, N_{0}, \ldots, N_{stop}-1$. The blue lines are ensemble averages based on $10,0000$ replicates. In all cases, $N_{0}=5$, $p_{g}=1$, and $p_{m} = 1$.}
\label{fig:figure7}
\end{center}
\end{figure}

\end{document}